\documentclass[%
 reprint,
 amsmath,
 amssymb,
 aps,
 pra,
 showpacs
]{revtex4-1}
\usepackage{float}
\usepackage{graphicx}% Include figure files
\usepackage{dcolumn}% Align table columns on decimal point
\usepackage{bm}% bold math
\usepackage{bbold}
\usepackage[export]{adjustbox}% http://ctan.org/pkg/adjustbox
\usepackage{amsmath,amsfonts,mathtools}
\usepackage{dsfont}
\usepackage{amsmath}
\usepackage{amssymb}
\usepackage{makeidx}
\usepackage{graphicx}
\usepackage{color}

\newcommand{\hide}[1]{}

\newcommand{\im}{\Im m\,}

\newcommand{\om}{\omega}

\newcommand{\lcase}{\left\{\begin{array}{ll}}
\newcommand{\rcase}{\end{array}\right.}
\newcommand{\ear}{\end{array}}
\newcommand{\bal}{\begin{align}}
\newcommand{\eal}{\end{align}}
\newcommand{\bma}{\begin{pmatrix}}
\newcommand{\ema}{\end{pmatrix}}
\newcommand{\beq}{\begin{equation}}
\newcommand{\eeq}{\end{equation}}
\newcommand{\bel}[1]{\begin{equation}\label{eq:#1}}
\newcommand{\eel}{\end{equation}}
\newcommand{\bea}{\begin{eqnarray}}
\newcommand{\eea}{\end{eqnarray}}
\newcommand{\beaNN}{\begin{eqnarray*}}
\newcommand{\eeaNN}{\end{eqnarray*}}

\newcounter{lecture}

\newcommand{\Ef}{\mathcal{E}}

\renewcommand{\hide}[1]{}

\renewcommand{\im}{\text{i}}
\newcommand{\hydroplus}{\text{H}_2^+}

\begin{document}

\preprint{APS/123-QED}

\title{Quantum Dynamics of $H_2^+$ in Orthogonal Two-Color Fields
%: Coherent Control of Fragmentation Channels and Anisotropic Energy Release
}
\author{Jinzhen Zhu}
\email{zhujinzhenlmu@gmail.com}
\affiliation{%
 Physics Department, Ludwig Maximilians Universit\"at, D-80333 Munich, Germany
}%
\affiliation{%
 Shanghai Artificial Intelligence Laboratory, 129 Longwen Road, Shanghai, China
}

\begin{abstract}
We present full-dimensional quantum simulations of $H_2^+$ dissociative ionization driven by strong orthogonal laser fields.
We consider equal-frequency orthogonal components, which generate elliptical or circular polarization depending on their relative phase and amplitude, as well as orthogonal $800$- and $400$-nm two-color fields.
These two-dimensional fields strongly modify the fragmentation dynamics. Most notably, we identify a high-energy peak in the proton kinetic-energy-release (KER) spectrum at approximately $4-5$ eV that is absent from the corresponding single-color, linearly polarized calculations.
The yield of this peak can be coherently controlled by varying the relative carrier-envelope phase of the perpendicular field component.
The perpendicular field also disrupts the clear electron--proton energy-sharing pattern observed in the main $3-3.5$ eV dissociation channel, indicating more complex multichannel dynamics.
Time-dependent state projections and calculations initiated from individual excited states attribute the additional peak to laser-induced vibrational excitation of $H_2^+$.
Furthermore, the perpendicular field rotates the fragment angular distributions, causing the most probable proton and electron emission directions to deviate substantially from the principal $z$ axis.
These findings demonstrate that the spatial and temporal geometry of orthogonal laser fields provides an additional degree of freedom for controlling ultrafast electron--nuclear dynamics.
\end{abstract}

\pacs{32.80.-t,32.80.Rm,32.80.Fb}
% 32.80.-t Photoionization and excitation
% 32.80.Rm Multiphoton ionization and excitation to highly excited states
% 32.80.Fb Photoionization of atoms and ions 
\maketitle

\section{\label{sec:intro}Introduction}
The hydrogen molecular ion $H_2^+$, the simplest molecule containing one electron and two protons, is a benchmark system for studying molecular dynamics in intense laser fields \cite{Krausz2009,Bucksbaum1990}. Its principal fragmentation mechanisms, including dissociative ionization (DI), above-threshold dissociation (ATD), and bond softening (BS), have been investigated extensively in both experiments and theory \cite{Giusti-Suzor1995,Odenweller2011,Odenweller2014,Wu2013,Gong2016}. These studies have shown that strong laser fields reshape molecular potential-energy surfaces and thereby control both the fragment kinetic energies and the direction of bond breaking \cite{Pavicic2005,Madsen2012,Scrinzi2012}. A characteristic example is the KER peak near $3-4$ eV that is frequently observed when $H_2^+$ dissociates in strong infrared fields \cite{Bucksbaum1990,Wu2013,Odenweller2014,Gong2016}. The initial vibrational state also strongly affects molecular excitation, ionization, and fragmentation \cite{Kulander1996}. In particular, fragmentation from excited vibrational states can produce additional features in the KER spectrum \cite{Zhou2005}, and excited-state populations can be essential for interpreting the structure of the joint energy spectrum (JES) \cite{Yue2014}.
\par
More complex laser fields provide additional control over these ultrafast processes. Tailoring the polarization and combining multiple frequencies introduce new degrees of freedom into the laser--molecule interaction \cite{Peng2015}. Circularly and elliptically polarized fields, for example, can suppress electron recollision with the parent ion and thereby reveal signatures of direct ionization and below-threshold dissociation (BTD) that may be obscured by interference effects in linearly polarized fields \cite{Znakovskaya2012}. Recent experiments have used different polarization states to measure molecular-frame photoelectron angular distributions (PADs), providing detailed information about molecular electron dynamics \cite{Granados2024,Wang2025a}. Two-color fields further enable coherent control through their relative phase, intensity, frequency, and polarization \cite{Charron1993,Wang2025}. Such fields can steer electron localization, induce asymmetric bond breaking, and modify dissociation probabilities \cite{Ray2009,Ma2025,Yue2013,Yue2014}. Nevertheless, full-dimensional quantum simulations of phase-dependent fragmentation in orthogonal fields remain challenging because the coupled electron--nuclear motion must be represented in six spatial dimensions \cite{Yue2013,Yue2014,Ranitovic2014}.
\par
Here, we present full-dimensional simulations of $H_2^+$ dissociative ionization in orthogonal laser fields. A primary $800$-nm component is polarized along the molecular $z$ axis, while a second $800$- or $400$-nm component is polarized along the perpendicular $x$ axis. We solve the TDSE with the tRecX code \cite{Scrinzi2010,Tao2012}. Related implementations have been applied to double ionization of helium \cite{Scrinzi2012,Zielinski2016,Zhu2020}, molecular single ionization \cite{Majety2015,Majety2015e,Majety2015d,Majety2015c,Majety2015g,Chundayil2024}, and full-dimensional $H_2^+$ dynamics in linearly polarized fields \cite{Zhu2020b,Zhu2021}. We identify an additional proton KER peak at approximately $4-5$ eV that is absent from the corresponding single-color calculations. Its yield varies strongly with the relative phase of the perpendicular component, demonstrating coherent control of this fragmentation channel. State-projection analysis and calculations initiated from individual excited states associate the peak with laser-induced vibrational excitation. These results show that orthogonal fields can manipulate correlated electron--nuclear dynamics through both their spatial geometry and their relative phase.
\section{Methods}
Atomic units, with $\hbar=e^2=m_e=4\pi\epsilon_0\equiv1$, are used unless stated otherwise.
We employ spherical coordinates centered at the midpoint between the two protons.
Rather than using the internuclear vector $\vec{R}$ as a coordinate~\cite{Yue2013,Yue2014,Madsen2012}, we place the protons at $\vec{r_1}$ and $-\vec{r_1}$ and denote the electron coordinate by $\vec{r_2}$.
The proton mass is denoted by $M=1836$ a.u.
\subsection{Hamiltonian}
The total Hamiltonian is the sum of the electron--proton interaction $H_{EP}$ and two one-particle Hamiltonians,
\begin{equation}\label{eq:HamiltonianH2PlusFull}
 H=H_{B}=H^{(+)}\otimes \mathds{1}+\mathds{1}\otimes H^{(-)}+H_{EP},
\end{equation}
where $\mathds{1}$ is the identity operator, $H^{(+)}$ is the nuclear Hamiltonian, and $H^{(-)}$ is the electronic Hamiltonian.
We denote the full Hamiltonian in the bound region by $H_B$.
After the coordinate transformation, the electronic Hamiltonian is
\begin{equation}
 H^{(-)}=-\frac{\Delta}{2m}-\im\beta\vec{A}(t)\cdot\vec{\triangledown },
\end{equation}
and the nuclear Hamiltonian is
\begin{equation}
  H^{(+)}=-\frac{\Delta}{4M}+\frac{1}{2r},
\end{equation}
where $m=\frac{2M}{2M+1}\approx1$ is the reduced electron mass and $\beta=\frac{1+M}{M}\approx1$.
The electron--proton interaction is
\begin{equation}
 H_{EP}=-\frac{1}{|\vec{r_1}+\vec{r_2}|}-\frac{1}{|\vec{r_1}-\vec{r_2}|}.
\end{equation}
\subsection{tSurff for dissociative ionization}
We calculate the JES using the tSurff method described in Refs.~\cite{Zhu2020b,Zhu2021}. The essential elements are summarized here for completeness.
\par
Within the tSurff approximation, all particle interactions are neglected beyond sufficiently large radii $R_c^{(+/-)}$. The corresponding asymptotic Hamiltonians are $H_V^{(+)}=-\frac{\Delta}{4M}$ for the nuclei and $H_V^{(-)}=-\frac{\Delta}{2m}-\im\beta\vec{A}(t)\cdot\vec{\triangledown}$ for the electron.
The nuclear scattering states satisfying $\im\partial_t \chi_{\vec{k_1}}(\vec{r_1})=H_V^{(+)}\chi_{\vec{k_1}}(\vec{r_1})$ are
\begin{equation}
  \chi_{\vec{k_1}}(\vec{r_1})=\frac{1}{(2\pi)^{3/2} }\exp(-\im\int_{t_0}^t\frac{k_1^2}{4M}d\tau)\exp(-\im\vec{k_1}\vec{r_1}),
\end{equation}
and the electronic scattering states satisfying $\im\partial_t \chi_{\vec{k_2}}(\vec{r_2})=H_V^{(-)}\chi_{\vec{k_2}}(\vec{r_2})$ are
\begin{equation}
 \chi_{\vec{k_2}}(\vec{r_2})=\frac{1}{(2\pi)^{3/2} }\exp(-\im\int_{t_0}^t \frac{k_2^2}{2m}-\im\beta\vec{A}(\tau)\cdot\vec{\triangledown}d\tau)\exp(-\im\vec{k_2}\vec{r_2}),
\end{equation}
where the laser field begins at $t_0$, while $\vec{k_1}$ and $\vec{k_2}$ denote the nuclear and electronic momenta, respectively.
\par
\begin{figure}
\centering
\includegraphics[width=0.4\textwidth]{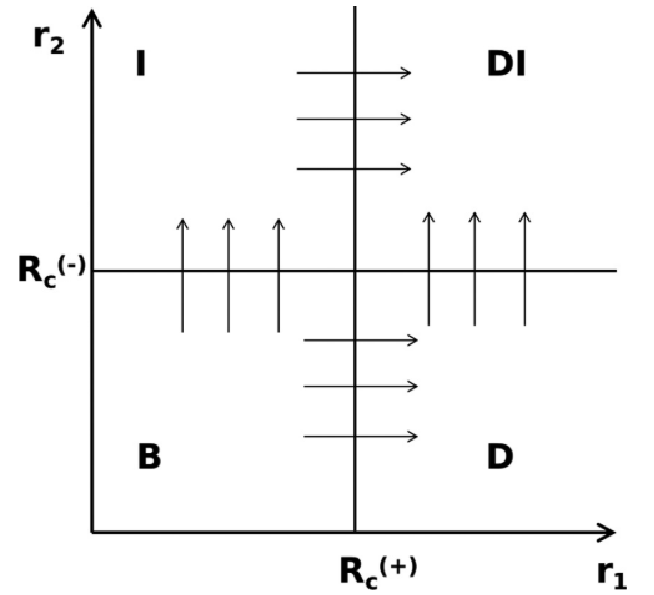}
\caption{Partition of configuration space used for the tSurff propagation of dissociative ionization. $B$ denotes the bound region. In $D$, the nuclei have crossed $R_c^{(+)}$ while the electron remains inside $R_c^{(-)}$; in $I$, the electron has crossed $R_c^{(-)}$ while the nuclei remain inside $R_c^{(+)}$. In the $DI$ region, both the nuclei and the electron have crossed their respective tSurff surfaces. Here, $R_c^{(+)}$ and $R_c^{(-)}$ are the tSurff radii associated with $r_1=|\vec{r_1}|$ and $r_2=|\vec{r_2}|$, respectively.
}
\label{fig:H2PlusRegions}
\end{figure}

The tSurff surfaces divide configuration space into four regions, denoted by $B$, $I$, $D$, and $DI$ in Fig.~\ref{fig:H2PlusRegions}. The bound region $B$ retains the full Hamiltonian of Eq.~(\ref{eq:HamiltonianH2PlusFull}). In the dissociation and ionization regions, propagation is governed by the single-particle Hamiltonians
\begin{equation}
 H_{D}(\vec{r_2},t)=H_{V}^{(-)}(\vec{r_2},t) = -\frac{\Delta}{2m}-\im\beta\vec{A}(t)\cdot\vec{\triangledown}
\end{equation}
and
\begin{equation}
 H_{I}(\vec{r_1},t) = -\frac{\Delta}{4M}+\frac{1}{2r_1},
\end{equation}
respectively, while the $DI$ contribution is obtained by time integration of the outgoing flux. This partition was introduced for the double ionization of helium in Ref.~\cite{Scrinzi2012} and subsequently applied to a two-dimensional model of $\hydroplus$ in Ref.~\cite{Yue2013}.
\par
For a sufficiently long propagation time $T$, we assume that the asymptotic electronic and nuclear scattering states are disentangled.
Introducing the step functions
\begin{equation}
 \Theta_{1/2}(R_c)=\left\{\begin{matrix}
0\;, r_{1/2}< R_c^{(+/-)} \\ 
1\;, r_{1/2}\geq  R_c^{(+/-)},
\end{matrix}\right.
\end{equation}
the unbound spectra can be written as
\begin{equation}\label{eq:finalSpectrum}
 P(\vec{k_1},\vec{k_2})=P(\phi_1,\theta_1,k_1,\phi_2,\theta_2,k_2)=\left | b(\vec{k_1},\vec{k_2},T) \right |^2.
\end{equation}
The scattering amplitude $b(\vec{k_1},\vec{k_2},T)$ is
\begin{equation}\label{eq:integralAmplitudes}
b(\vec{k_1},\vec{k_2},T)=\int_{-\infty}^{T}[F(\vec{k_1},\vec{k_2},t)+\bar{F}(\vec{k_1},\vec{k_2},t)] dt
\end{equation}
with two source terms,
\begin{equation}\label{eq:Fk1k2H2Plus}
 F(\vec{k_1},\vec{k_2},t)= \langle \chi_{\vec{k_2}}(\vec{r_2},t)\left |[H_V^{(-)}(\vec{r_2},t), \Theta_2(R_c)] \right | \varphi_{\vec{k_1}}(\vec{r_2},t) \rangle
\end{equation}
and 
\begin{equation}\label{eq:Fk1k2H2PlusBar}
 \bar{F}(\vec{k_1},\vec{k_2},t)= \langle \chi_{\vec{k_1}}(\vec{r_1},t)\left | [H_V^{(+)}(\vec{r_1},t), \Theta_1(R_c)] \right | \varphi_{\vec{k_2}}(\vec{r_1},t) \rangle.
\end{equation}
The single-particle wave functions $\varphi_{\vec{k_1}}(\vec{r_2},t)$ and $\varphi_{\vec{k_2}}(\vec{r_1},t)$ satisfy
\begin{equation}\label{eq:unbound0}
 \im\frac{d}{dt}\varphi_{\vec{k_1} }(\vec{r_2},t)=H_D(\vec{r_2},t) \varphi_{\vec{k_1} }(\vec{r_2},t)-C_{\vec{k_1} }(\vec{r_2},t)
\end{equation}
and
\begin{equation}\label{eq:unbound1}
 \im\frac{d}{dt}\varphi_{\vec{k_2} }(\vec{r_1},t)=H_I(\vec{r_1},t)\varphi_{\vec{k_2} }(\vec{r_1},t)-C_{\vec{k_2} }(\vec{r_1},t).
\end{equation}
The source terms are projections of the full wave function onto the asymptotic solutions,
\begin{equation}\label{eq:source1}
 C_{\vec{k_1} }(\vec{r_2},t)=\int d\vec{r_1} \overline{\chi_{\vec{k_1}}(\vec{r_1},t)}[H_V^{(+)}(\vec{r_1},t),\Theta_1(R_c)]\psi(\vec{r_1},\vec{r_2},t)
\end{equation}
and
\begin{equation}\label{eq:source2}
 C_{\vec{k_2} }(\vec{r_1},t)=\int d\vec{r_2} \overline{\chi_{\vec{k_2}}(\vec{r_2},t)}[H_V^{(-)}(\vec{r_2},t),\Theta_2(R_c)]\psi(\vec{r_1},\vec{r_2},t),
\end{equation}
with zero initial values. We use equal tSurff radii, $R_c^{(+)}=R_c^{(-)}$. Beyond either surface, all Coulomb interactions involving the outgoing particle are neglected. Previous studies have shown that the resulting spectra become independent of $R_c$ when the asymptotic Hamiltonian is used consistently and the wave function is propagated sufficiently long after the pulse~\cite{Zielinski2016,Scrinzi2012}. Infinite-range exterior complex scaling (irECS) is used to absorb outgoing flux~\cite{Scrinzi2010}.
\section{Numerical results}
We use the angular basis $0 \leq m_{1/2} \leq 2$ and $0 \leq l_{1/2} \leq 8$, consistent with the parameters employed in our previous full-dimensional study of linearly polarized ionization~\cite{Zhu2020b}. The enlarged basis is required by the reduced symmetry of the present orthogonal-field calculations. For comparison, six-dimensional double-emission calculations for helium can exploit axial symmetry and converge with $m_{1/2}=0$ and $0 \leq l_{1/2} \leq 2$~\cite{Zielinski2016}.
The cutoff radii are $R_c^{(+)}=R_c^{(-)}=12.5$ a.u., corresponding to a maximum internuclear separation of $R=25$ a.u. and consistent with previous benchmarks~\cite{Yue2013}. The wave function is propagated sufficiently long after the pulse to resolve low-energy outgoing contributions. A complete convergence study at the reduced symmetry of the orthogonal field would require substantially greater memory and computational time. Nevertheless, calculations with coarser angular representations reproduce the secondary peak at the same KER. Together with the use of parameters established in the previous linearly polarized study, this indicates that the peak position and the qualitative trends discussed below are numerically stable, although the precise relative yields may retain some basis dependence.
\par
Including nuclear kinetic energy, we obtain a field-free ground-state energy of $E_0=-0.592$ a.u. and an equilibrium internuclear distance of $2.05$ a.u.
When nuclear kinetic energy is excluded, the ground-state energy is $-0.597$ a.u., in agreement to three decimal places with the fixed-nuclei quantum-chemistry result of Ref.~\cite{Bressanini1997}. The corresponding equilibrium distance is $1.997$ a.u., also agreeing to three decimal places with the accurate value reported in Ref.~\cite{Schaad1970}.
\subsection{Laser pulses}
\label{sec:pulses}
The peak intensities of the $z$- and $x$-polarized components are $I_z=\Ef_{z,0}^2$ and $I_x=\Ef_{x,0}^2$, respectively, in atomic units. The total field is the superposition of the two orthogonal components,
\begin{equation}
    \vec{\Ef}(t)=\Ef_z(t)\hat{z}+\Ef_x(t)\hat{x}
\end{equation}
The components are obtained from their vector potentials through $\Ef_z(t)=-\partial_t A_z(t)$ and $\Ef_x(t)=-\partial_t A_x(t)$, where
\begin{equation}
\begin{split}
A_z(t)=\frac{\Ef_{z,0}}{\om_z} a_z(t)\sin(\om_z t+\phi_{z,CEP})\\
A_x(t)=\frac{\Ef_{z,0}}{\om_x} a_x(t)\sin(\om_x t+\phi_{x,CEP})\\
\end{split}
\end{equation}
The $z$ axis is aligned with the molecular axis, and the $x$ axis is perpendicular to it. For equal frequencies, envelopes, and amplitudes, a relative phase of $\pi/2$ or $3\pi/2$ produces circular polarization; other relative phases or unequal amplitudes produce elliptical polarization. When $\omega_x\neq\omega_z$, the field is an orthogonally polarized two-color field rather than a conventional circularly polarized field. We use either a $\cos^8$ envelope, $a(t)=[\cos(\frac{t}{n\tau})]^8$, or the flat-top envelope defined in Eq.~\ref{eq:flatTop}. Here, $n$ specifies the number of optical cycles associated with the full width at half maximum (FWHM).

\subsection{Joint energy spectra}
\begin{figure}
\centering
\includegraphics[width=0.23\textwidth,trim=1.2cm 0cm 1.5cm 0.5cm,clip]{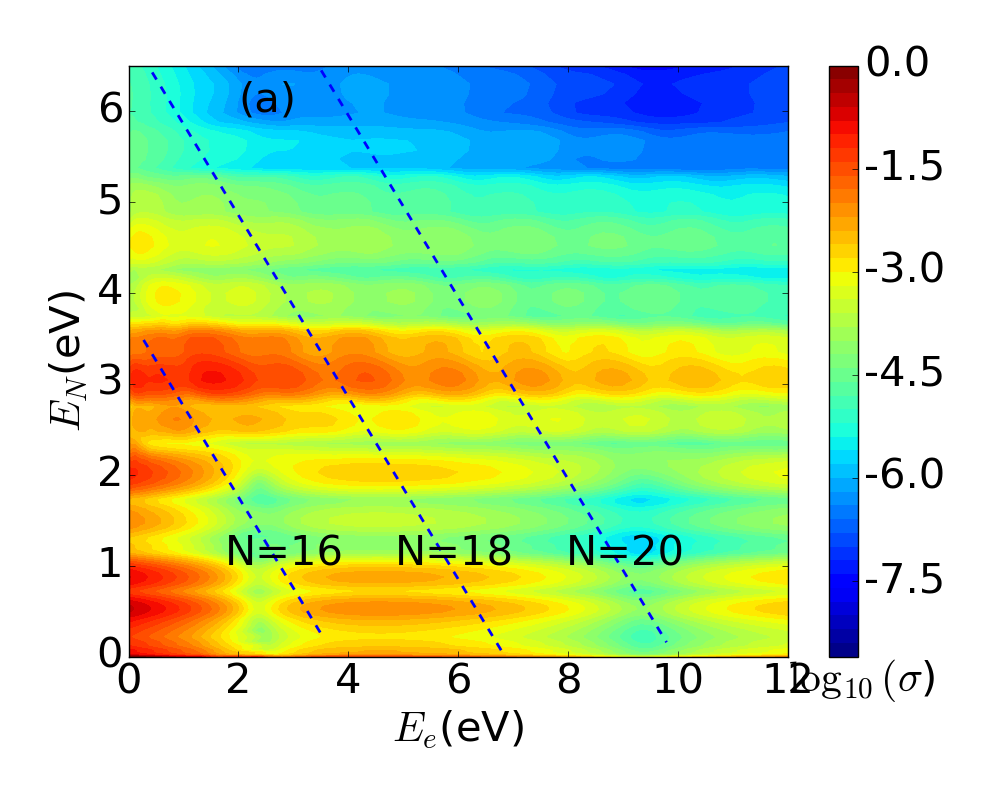}
\includegraphics[width=0.23\textwidth,trim=1.2cm 0cm 1.5cm 0.5cm,clip]{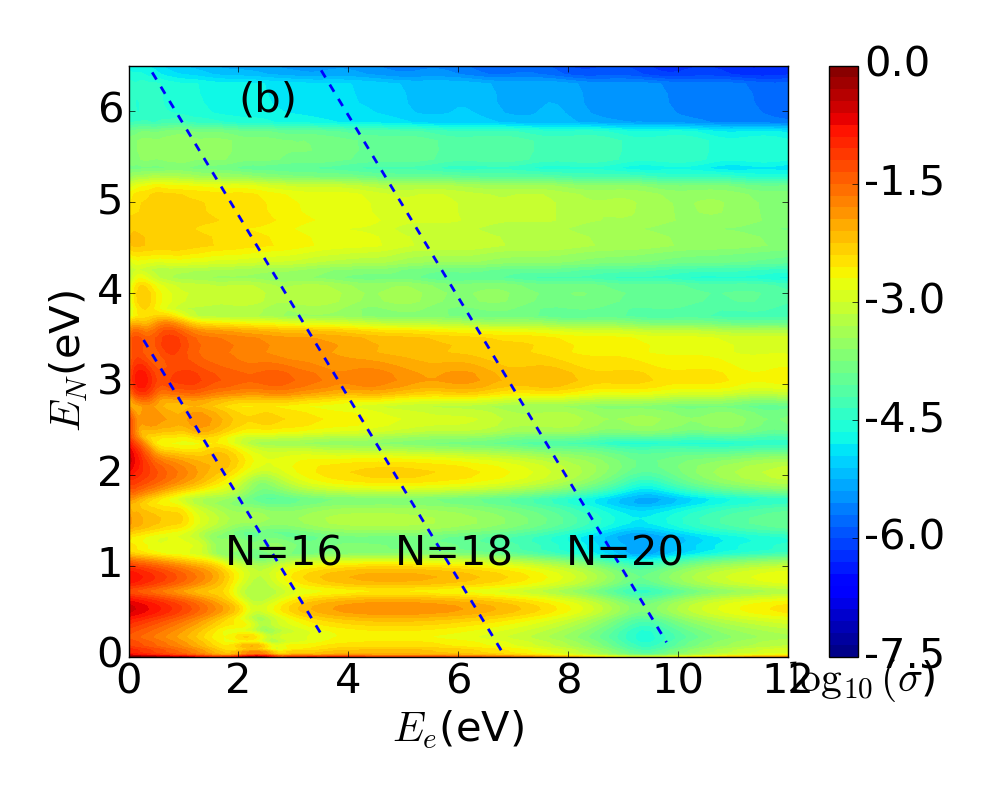}
\includegraphics[width=0.23\textwidth,trim=1.2cm 0cm 1.5cm 0.5cm,clip]{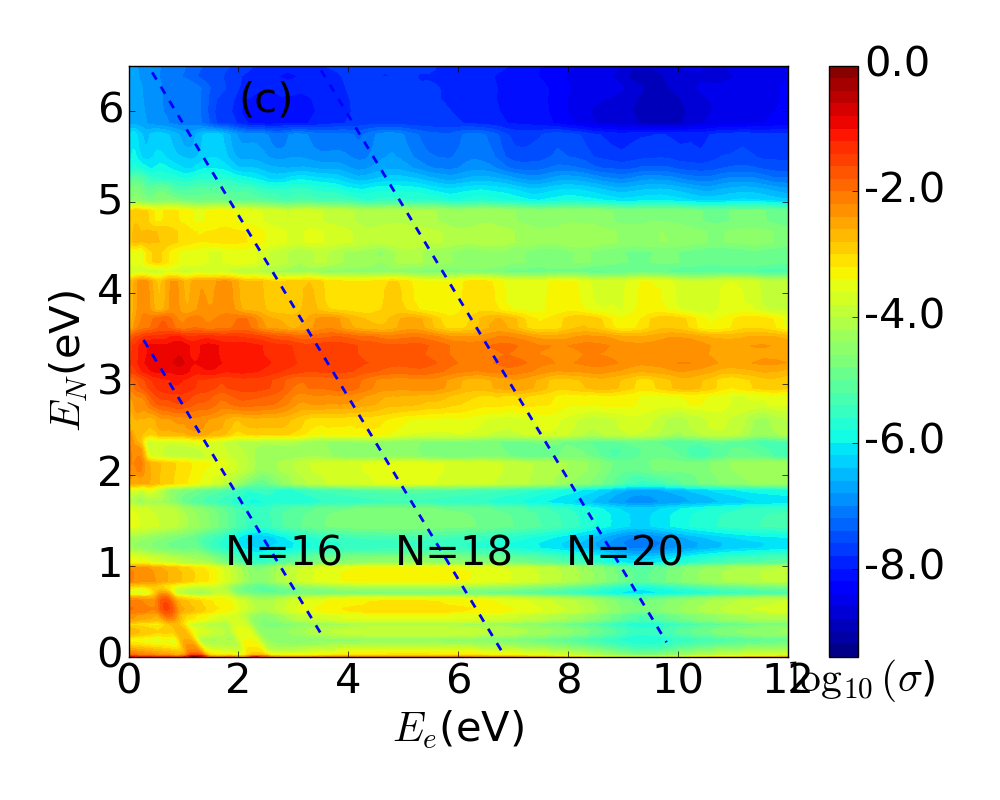}
\includegraphics[width=0.23\textwidth,trim=1.2cm 0cm 1.5cm 0.5cm,clip]{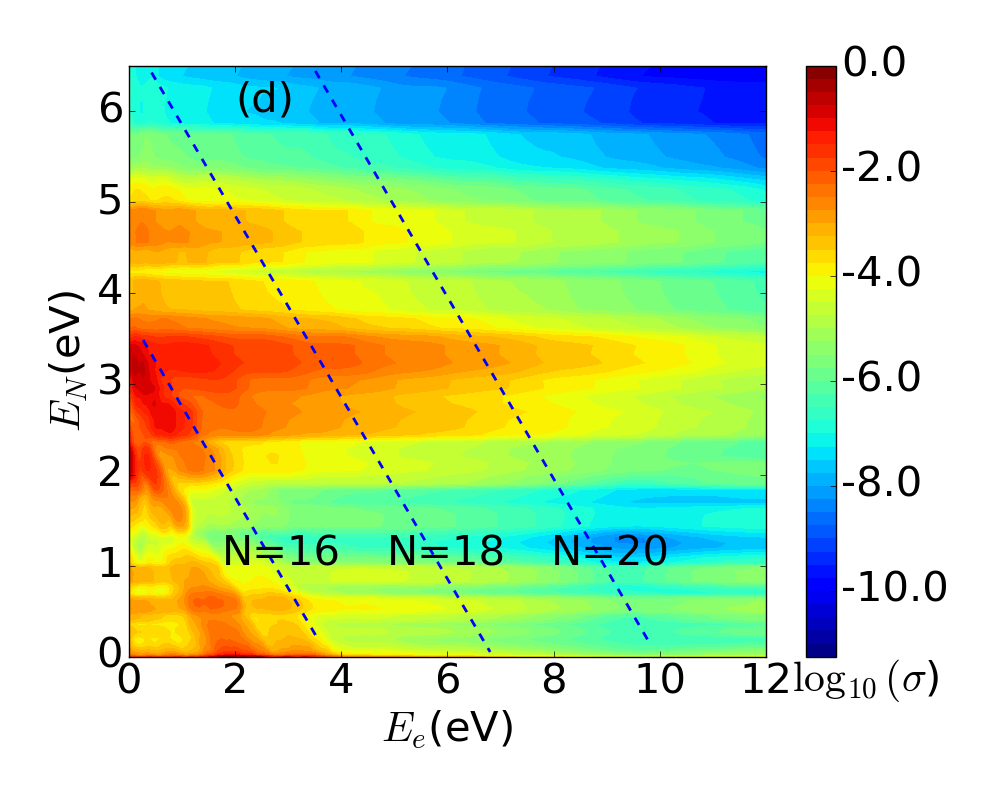}
\caption{JES of $\text{H}_2^+$ dissociative ionization driven by $800$-nm, $\cos^8$ pulses with a peak intensity of $8\times10^{13}\ \text{W/cm}^2$: (a) linearly polarized, FWHM $2$ optical cycles; (b) circularly polarized, FWHM $2$ optical cycles; (c) linearly polarized, FWHM $6$ optical cycles; and (d) circularly polarized, FWHM $6$ optical cycles. The dashed lines show the energy-sharing condition of Eq.~\ref{eq:energySharing} for absorption of $N$ photons.}
\label{fig:800nmLongShortCompare}
\end{figure}

The JES of the three outgoing particles is obtained by integrating Eq.~\ref{eq:finalSpectrum} over all angular coordinates,
\begin{equation}
\begin{split}
 \sigma(E_N,E_e)=&\int d\phi_1\int d\phi_2\int d\theta_1\sin\theta_1\int d\theta_2\sin\theta_2\\
 &P(\phi_1,\theta_1,\sqrt{4M\times E_N},\phi_2,\theta_2,\sqrt{2m\times E_{e} }),
 \end{split}
\end{equation}
where $E_N$ is the total kinetic energy of the two protons and $E_e$ is the electron kinetic energy. The quantity $\sigma(E_N,E_e)$ is shown in Fig.~\ref{fig:800nmLongShortCompare} and in the subsequent JES plots. The tilted lines overlaid on these spectra follow the energy-conservation condition

\begin{equation}\label{eq:energySharing}
  E_{N}+E_{e}=N\omega+E_0 -U_p
\end{equation}
with $U_p=\frac{A_{z,0}^2}{4m}$. The lines indicate the energy sharing expected after absorption of $N$ photons, where $E_0$ is the molecular ground-state energy. Structure aligned with these lines reflects correlated electron--nuclear emission, as also observed in our previous $400$-nm calculations~\cite{Zhu2020b,Zhu2021}. Unless stated otherwise, the spectra are normalized for comparison of their shapes and relative peak strengths; no conclusions are drawn from their absolute magnitudes.
\par
Figure~\ref{fig:800nmLongShortCompare} compares the JES obtained with short ($2$ optical cycles) and long ($6$ optical cycles) $800$-nm pulses under linear and circular polarization. A distinct energy-sharing pattern is visible in the linearly polarized cases, panels (a) and (c), and becomes sharper for the longer pulse. Under circular polarization, panels (b) and (d), this structure is less clearly resolved. In all four cases, the dominant proton KER lies near $2-4$ eV. An additional peak at approximately $4-5$ eV appears only in the circularly polarized calculations and is present for both pulse durations.

\par
To test the robustness and controllability of the additional peak, we performed calculations for several perpendicular $x$-polarized fields. The primary $z$-polarized pulse was fixed at $800$ nm, $8\times10^{13}\ \text{W/cm}^2$, a $\cos^8$ envelope with FWHM $2$ optical cycles, and carrier-envelope phase $\phi_{\text{CEO}}=0$. The frequency, intensity, and phase of the $x$-polarized control field were varied. We quantify the prominence of the additional channel by the ratio of its peak height to that of the main KER peak,
\begin{equation}
  H=\frac{\max_{4\leq E_N <5}\sigma(E_N)}{\max_{2\leq E_N< 4}\sigma(E_N)},\sigma(E_N)=\int\sigma(E_N,E_e)dE_e.
\end{equation}
This dimensionless measure compares spectral shapes independently of their overall normalization.

\par
\begin{figure}
\centering
\includegraphics[width=0.48\textwidth,trim=0cm 0cm 0cm 0cm,clip]{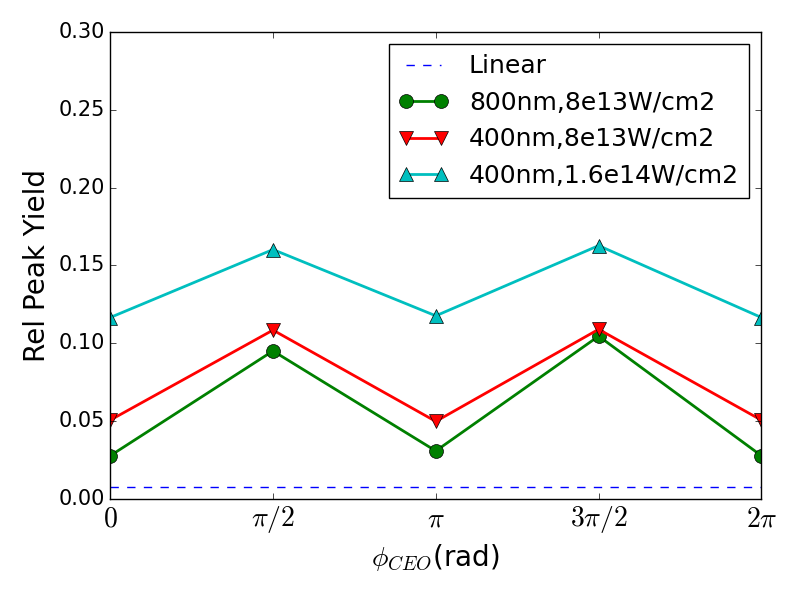}
\caption{Relative height of the secondary proton KER peak at $4-5$ eV with respect to the main peak at $2-4$ eV, plotted as a function of the relative phase of the perpendicular field. The $z$-polarized pulse is fixed at $800$ nm, $8\times10^{13}\ \text{W/cm}^2$, with a $\cos^8$ envelope of FWHM $2$ optical cycles. The dashed line is the single-color, linearly polarized reference. The control-field parameters are $800$ nm and $8\times10^{13}\ \text{W/cm}^2$ (green), $400$ nm and $8\times10^{13}\ \text{W/cm}^2$ (red), and $400$ nm and $1.6\times10^{14}\ \text{W/cm}^2$ (cyan).}
\label{fig:secondaryPeakHeight}
\end{figure}

As shown in Fig.~\ref{fig:secondaryPeakHeight}, the relative height of the $4-5$ eV peak depends strongly on the phase of the perpendicular field. Every orthogonal-field calculation produces a larger secondary contribution than the single-color, linearly polarized reference (dashed line). For the equal-frequency configuration, the enhancement is largest near $\phi_{\text{CEO}}=\pi/2$ and $3\pi/2$, where the field is circularly polarized, and smallest near $0$ and $\pi$, where the combined field is linearly polarized. Increasing the control-field intensity or reducing its wavelength further enhances the secondary contribution, as illustrated by the red and cyan curves. The corresponding JES are shown in Figs.~\ref{fig:normal400}, \ref{fig:high400}, and \ref{fig:800nsShort}.

These results may also help interpret the broader experimental KER distribution reported for circular polarization by Odenweller \textit{et al.}~\cite{Odenweller2014}. Relative to the linearly polarized result, the circularly polarized distribution retains more yield at higher KER and therefore appears flatter. Although the experiment did not resolve a distinct peak at $4-5$ eV, the secondary channel found here would enhance this high-energy region and may contribute to the observed broadening. This comparison remains qualitative because the experimental intensity and pulse duration differ from those used in the present calculations.
% \begin{figure}
% \centering
% \includegraphics[width=0.48\textwidth,trim=4cm 0cm 1.5cm 0.5cm,clip]{figs/sum-800nm.png}
% \includegraphics[width=0.48\textwidth,trim=4cm 0cm 1.5cm 0.5cm,clip]{figs/sum-400nm.png}
% \includegraphics[width=0.48\textwidth,trim=4cm 0cm 1.5cm 0.5cm,clip]{figs/sum-400nmhigh.png}
% \caption{z:800nm 8e13 2 cycles, x:800nm 8e13 2 cycles. fig:normal400 fig:high400 fig:800nsShort
% }
% \label{fig:800nsShortCompare}
% \end{figure}

\par
\subsection{Population analysis}
\begin{figure}
\centering
\includegraphics[width=0.23\textwidth,clip]{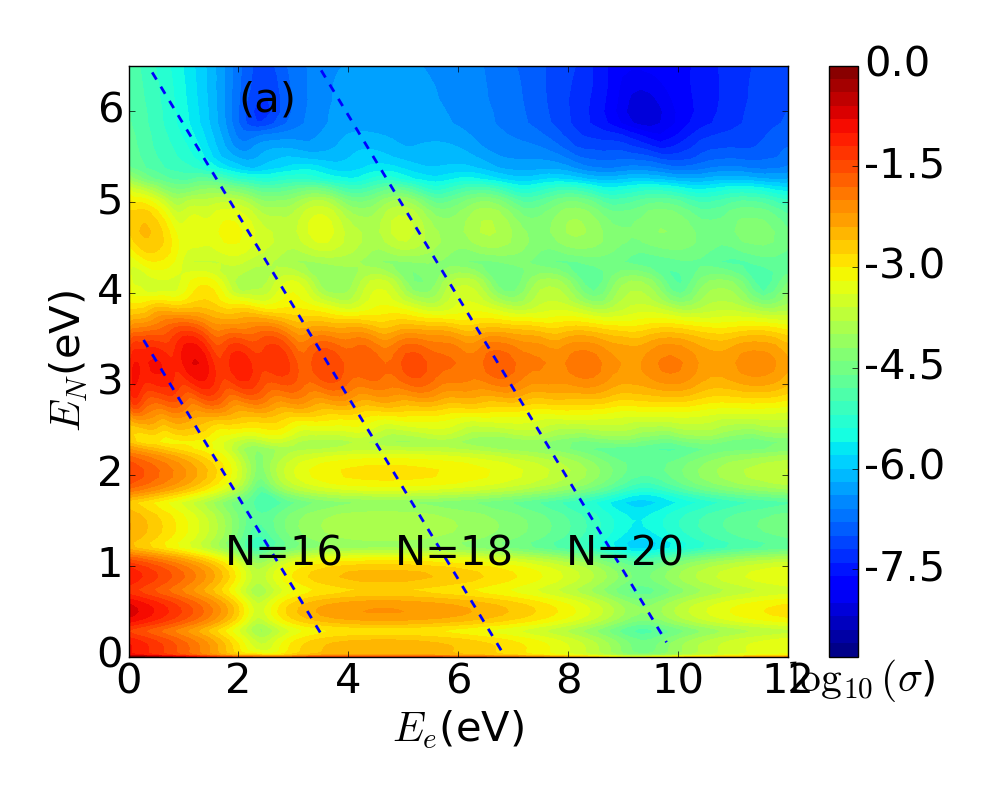}
\includegraphics[width=0.23\textwidth,clip]{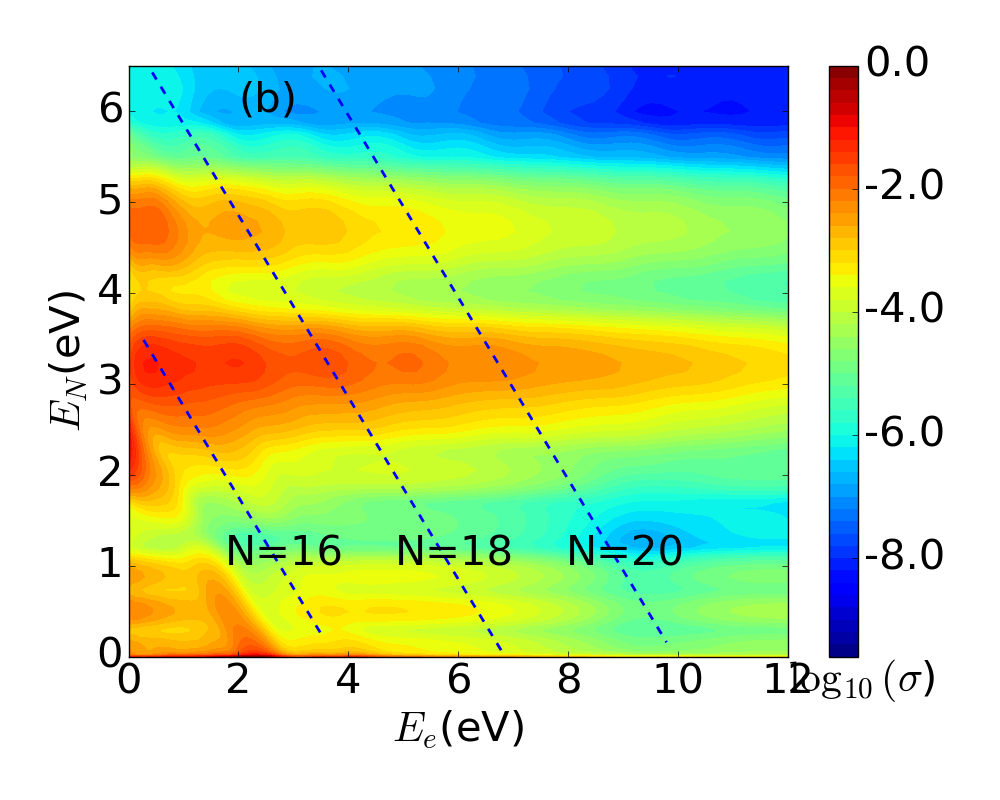}
\caption{JES of $\text{H}_2^+$ dissociative ionization driven by an $800$-nm flat-top pulse with FWHM $6$ optical cycles and peak intensity $8\times10^{13}\ \text{W/cm}^2$: (a) linear polarization and (b) circular polarization. The dashed lines show the energy-sharing condition of Eq.~\ref{eq:energySharing}.}
\label{fig:flatTopLinearCircular}
\end{figure}

To investigate the origin of the additional $4-5$ eV proton KER peak and exclude an artifact of the $\cos^8$ envelope, we repeated the calculation with a flat-top pulse. The pulse has a wavelength of $800$ nm, a peak intensity of $8\times10^{13}\ \text{W/cm}^2$, and a FWHM of $6$ optical cycles. As shown in Fig.~\ref{fig:flatTopLinearCircular}, circular polarization again produces the additional peak, whereas the linearly polarized calculation retains a clear energy-sharing structure.

\par
\begin{figure}
\centering
\includegraphics[width=0.46\textwidth,clip]{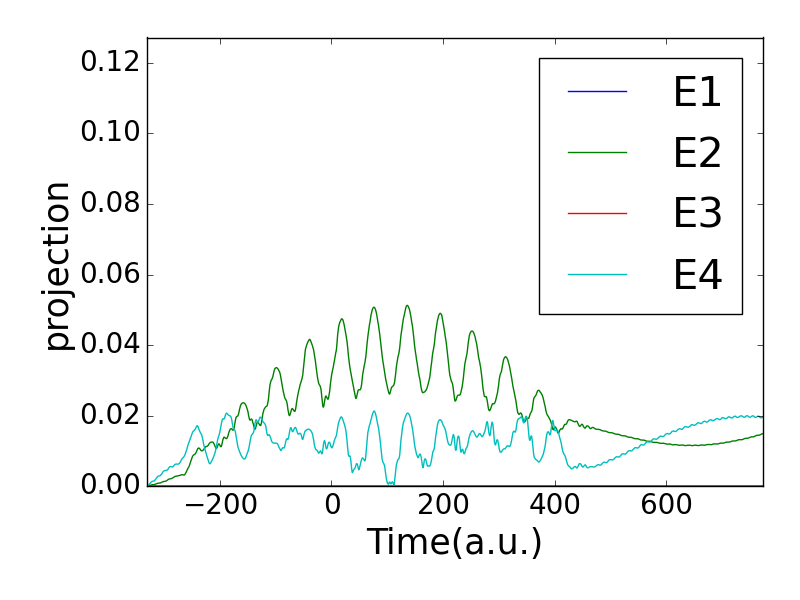}
\includegraphics[width=0.46\textwidth,clip]{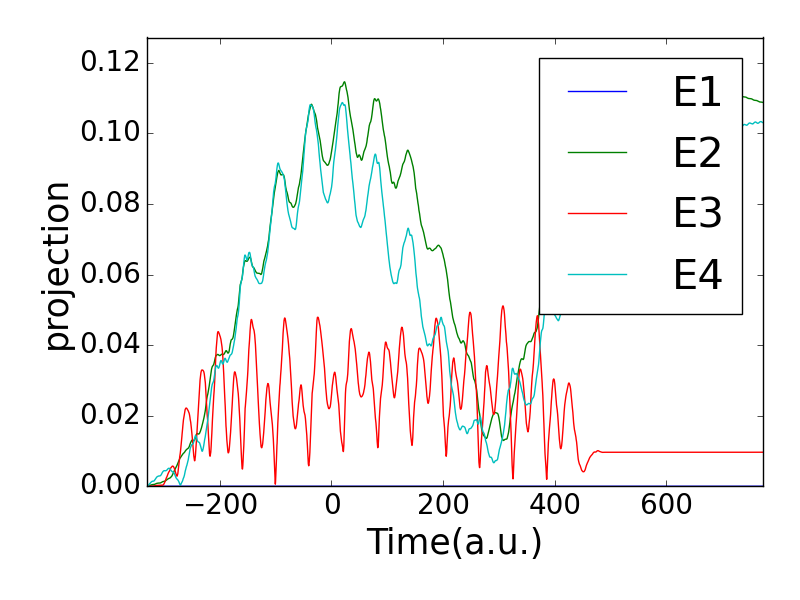}
\caption{Time-dependent magnitudes of the projection amplitudes, $|\langle E_v|\psi(t)\rangle|$, for the first four excited eigenstates of $\text{H}_2^+$. These quantities are used as relative indicators of laser-induced excitation. The molecule is driven by an $800$-nm flat-top pulse with peak intensity $8\times10^{13}\ \text{W/cm}^2$ and FWHM $6$ optical cycles. The upper and lower panels show linear and circular polarization, respectively.}
\label{fig:flatTopPopulation}
\end{figure}
\begin{figure}
\centering
\includegraphics[width=0.23\textwidth,clip]{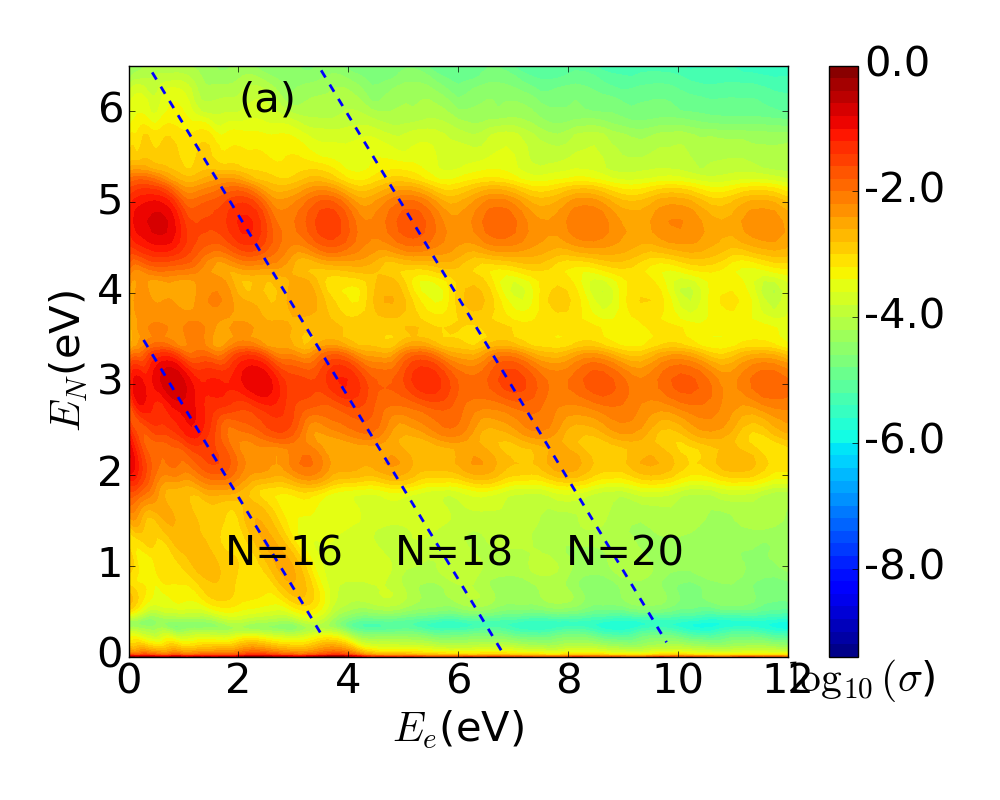}
\includegraphics[width=0.23\textwidth,clip]{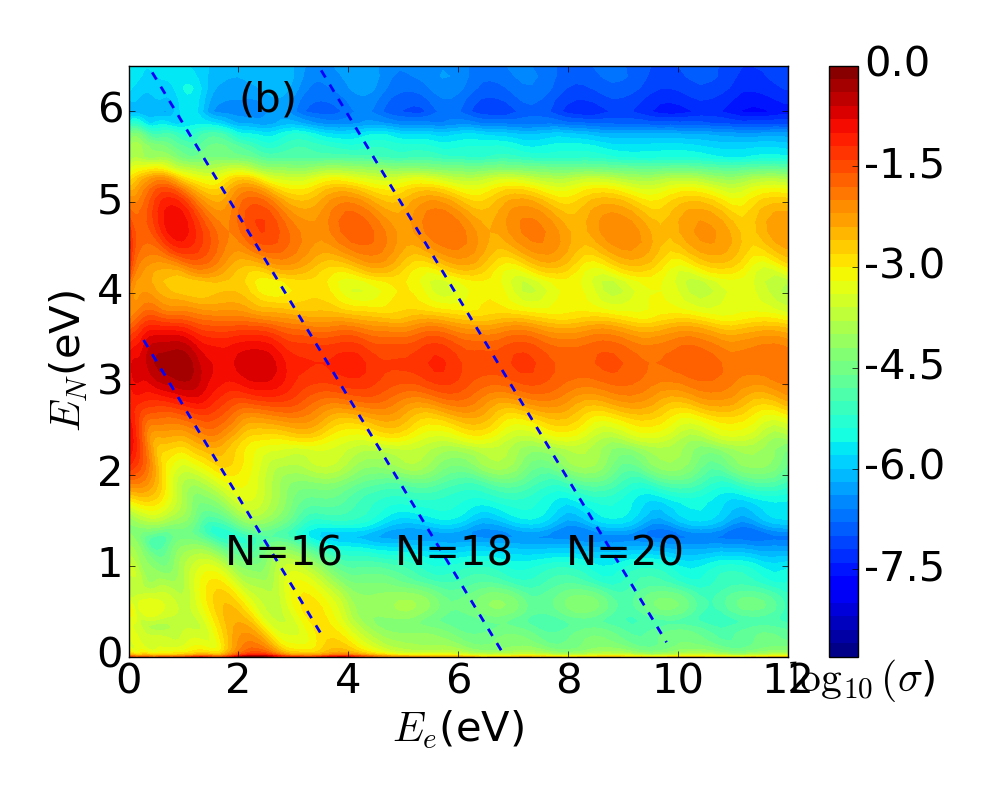}
\includegraphics[width=0.23\textwidth,clip]{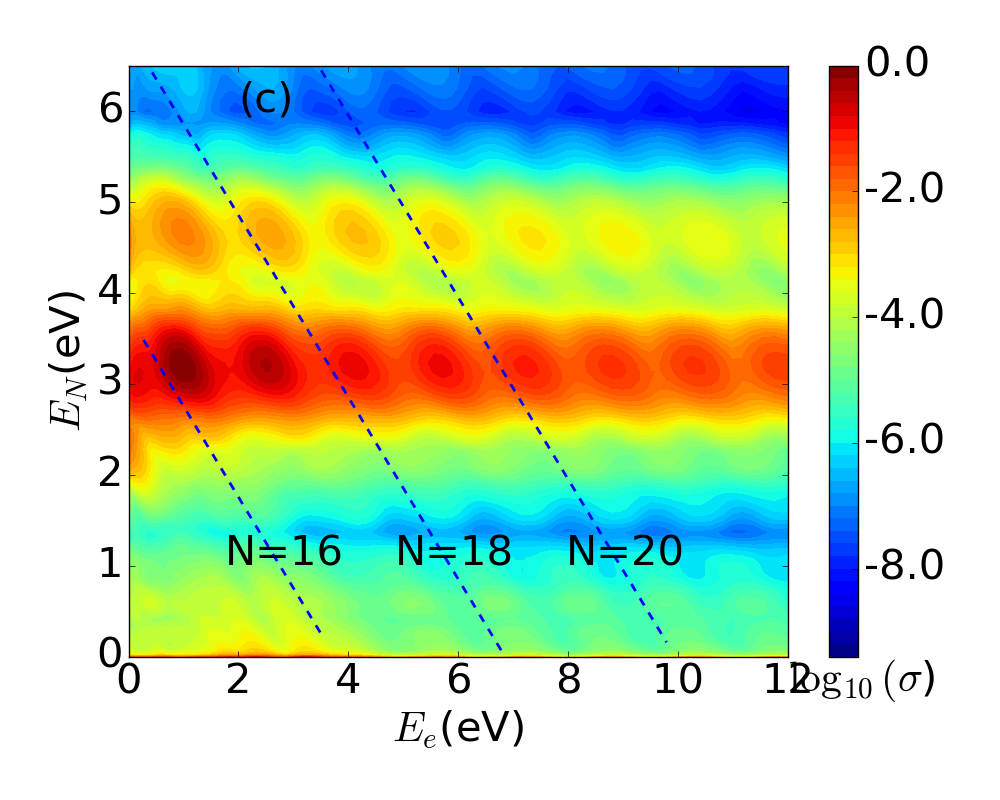}
\includegraphics[width=0.23\textwidth,clip]{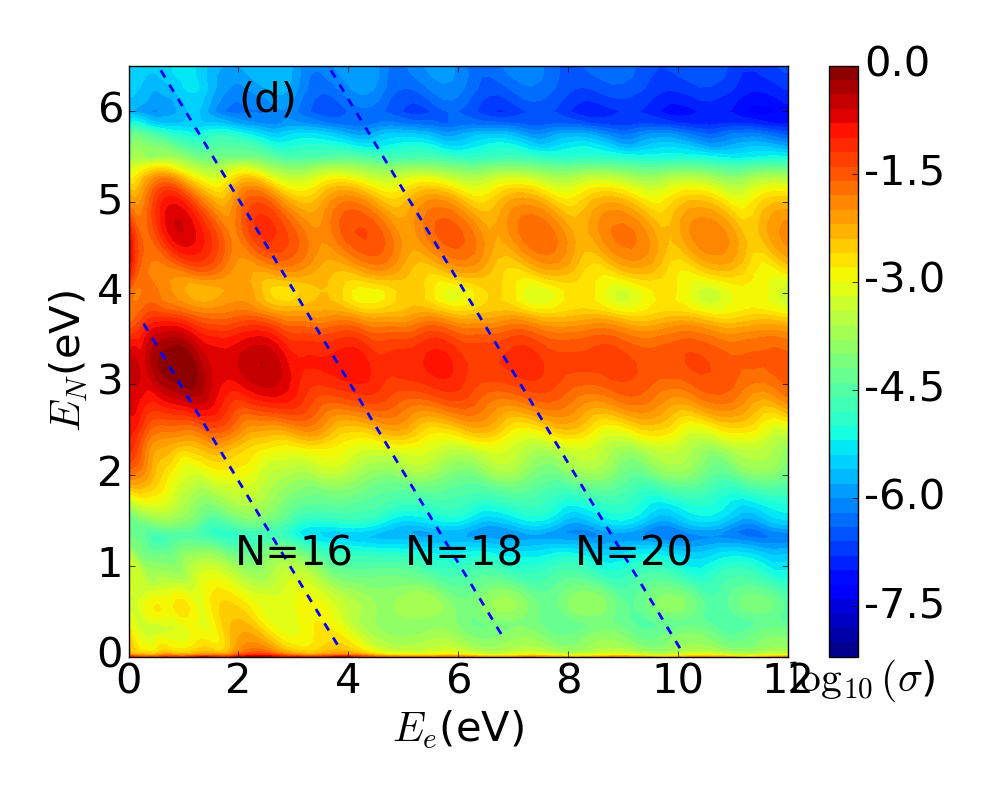}
\caption{JES obtained from calculations initiated in individual excited eigenstates: (a) $E_1$, (b) $E_2$, (c) $E_3$, and (d) $E_4$. A linearly polarized $800$-nm flat-top pulse is applied along the $z$ axis, with peak intensity $8\times10^{13}\ \text{W/cm}^2$ and FWHM $6$ optical cycles. The dashed lines show the energy-sharing condition of Eq.~\ref{eq:energySharing}.}
\label{fig:flatTopSpectrum}
\end{figure}

We calculated the first four excited eigenstates of $\text{H}_2^+$ and monitored the magnitudes of their projection amplitudes during linearly and circularly polarized pulses. Figure~\ref{fig:flatTopPopulation} shows that circular polarization produces substantially larger projections onto $E_2$ and $E_4$, indicating more efficient excitation of these states by the perpendicular field component. To determine whether these excited components can generate the secondary peak, we performed additional linearly polarized calculations initiated from each excited state rather than from the ground state. The resulting JES are shown in Fig.~\ref{fig:flatTopSpectrum}. Calculations initiated from $E_2$ and $E_4$ reproduce the $4-5$ eV peak with a strength comparable to that of the main peak near $3$ eV. This state-resolved test links the secondary channel to enhanced excitation of $E_2$ and $E_4$ and is consistent with earlier studies showing that vibrational excitation can strongly modify fragmentation spectra~\cite{Kulander1996,Zhou2005,Yue2014}.

Reference~\cite{Zhu2026Geometric} provides a complementary geometric picture for the sudden fragmentation of a single hydrogen-like radial state. In that model, a radial wave function $\psi_v(r)$ is mapped onto a distribution of local fragment energies,
\begin{equation}\label{eq:localFragmentEnergy}
 E_v(r)=-\frac{1}{2\mu}\frac{\nabla^2\psi_v(r)}{\psi_v(r)}+\frac{Q}{r},
\end{equation}
where $\mu$ is the effective radial mass, the first term is the local quantum kinetic energy, and $Q/r$ is the repulsive Coulomb energy. The corresponding spectrum is obtained by sampling $E_v(r)$ with the three-dimensional radial probability density,
\begin{equation}\label{eq:geometricEnergyDistribution}
 P_v(E)=\int_0^\infty 4\pi r^2|\psi_v(r)|^2
 \delta\!\left[E-E_v(r)\right]dr.
\end{equation}
The factor $4\pi r^2$ acts as a geometric filter that connects the spatial structure of the initial state to the KER distribution. Radial nodes generate rapid variations in the curvature term $\nabla^2\psi_v/\psi_v$ and can therefore redistribute spectral weight toward additional kinetic-energy features.

This radial single-state model is not evaluated directly for the present eigenstates, which contain multiple orbital components and nontrivial angular dependence. It should therefore be regarded as a qualitative analogy rather than a quantitative derivation of the $4-5$ eV peak. Nevertheless, it illustrates how the different nodal structures and spatial distributions of excited components can produce fragmentation energies that differ from those of the ground state. Together with the state-projection and excited-state calculations, this picture supports an excitation-assisted origin of the secondary channel. The state-resolved JES also display energy-sharing structures associated with both the main and secondary peaks.

\subsection{Angular distribution}
We finally examine how the perpendicular field modifies the proton and electron angular distributions. The distributions are projected onto the $xz$ plane, perpendicular to the laser-propagation direction $y$, by integrating over the remaining momentum coordinates as in Refs.~\cite{Zhu2020b,Zhu2021}. Starting from the six-dimensional probability distribution $P(\vec{k_1},\vec{k_2})$ of Eq.~\ref{eq:finalSpectrum}, the proton distribution is
\begin{equation}\label{eq:proton-angle}
 p_{N}(\theta, E)=\left\{\begin{matrix}
\int d \vec{k_2} P(0,\theta,\sqrt{8M E},\vec{k_2}), 0\leq\theta< \pi \\ 
\int d \vec{k_2} P(\pi,2\pi-\theta,\sqrt{8M E},\vec{k_2}),\pi\leq\theta<2\pi
\end{matrix}\right\}
\end{equation}
and the electron distribution is
\begin{equation}\label{eq:elctron-angle}
 p_{e}(\theta, E)=\left\{\begin{matrix}
\int d \vec{k_1} P(\vec{k_1},0,\theta,\sqrt{2m E}), 0\leq\theta< \pi \\ 
\int d \vec{k_1} P(\vec{k_1},\pi,2\pi-\theta,\sqrt{2m E}),\pi\leq\theta<2\pi
\end{matrix}\right\}
\end{equation}
Here, $E$ denotes the kinetic energy of an individual proton or electron rather than the total nuclear KER.

\par
\begin{figure}
\centering
\includegraphics[width=0.5\textwidth,clip]{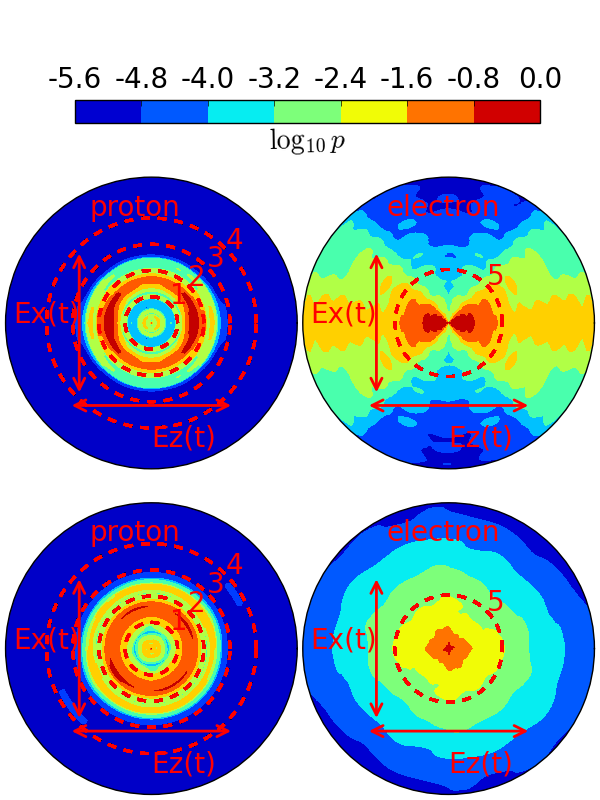}
\caption{
Logarithmic angular probability distributions for $\text{H}_2^+$ dissociative ionization driven by $800$-nm, $\cos^8$ pulses with peak intensity $8\times10^{13}\ \text{W/cm}^2$ and FWHM $6$ optical cycles. The upper and lower rows correspond to linear and circular polarization, respectively; the left and right columns show protons and electrons. The radial coordinate $E_{1/2}$ is the individual-fragment kinetic energy. Arrows indicate the directions of $E_z(t)$ and $E_x(t)$. Each distribution is normalized to its maximum.
}
\label{fig:800nmLongAngle}
\end{figure}

Figure~\ref{fig:800nmLongAngle} compares the angular distributions produced by linearly and circularly polarized $800$-nm, $\cos^8$ pulses with peak intensity $8\times10^{13}\ \text{W/cm}^2$ and FWHM $6$ optical cycles. Under linear polarization, both protons and electrons are emitted preferentially along the $z$ axis. Circular polarization rotates and broadens the distributions, producing substantial emission toward both positive and negative $x$. No clear preference between the two $x$ directions remains for either particle. Thus, for the long pulses considered here, the relative phase has little influence on directional asymmetry.
\par
\begin{figure}
\centering
\includegraphics[width=0.40\textwidth,clip]{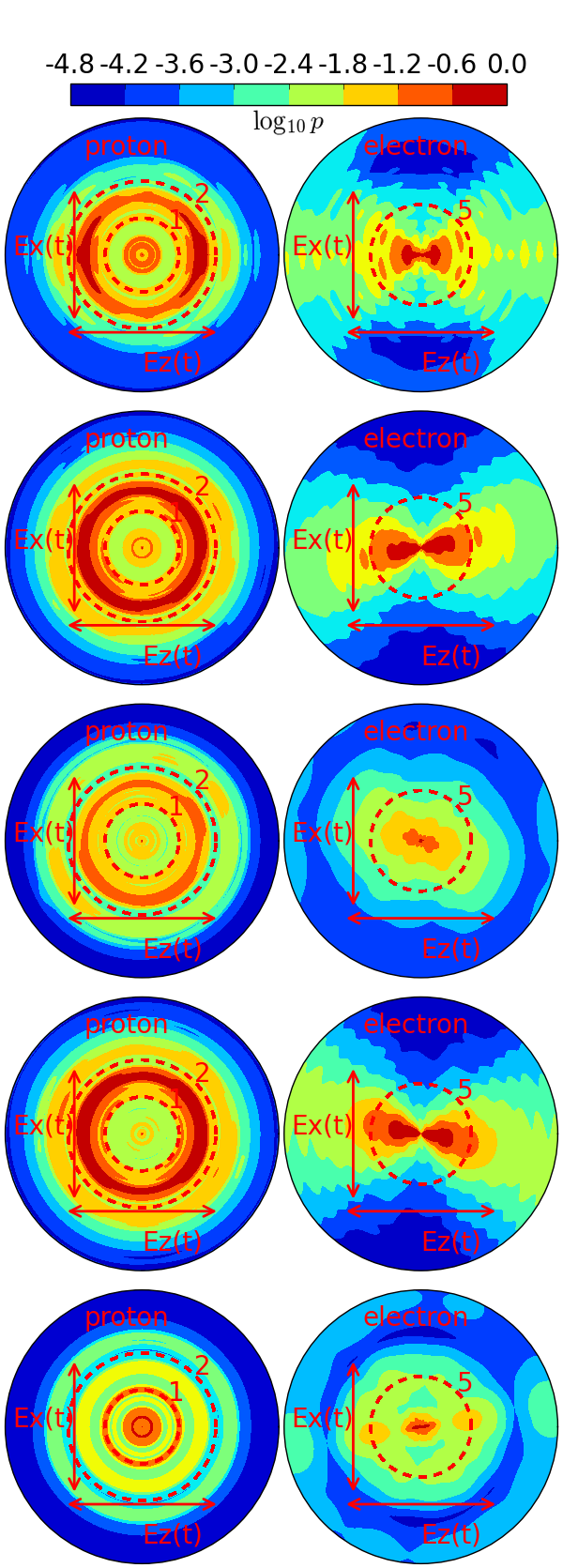}
\caption{Logarithmic fragment angular distributions for $\cos^8$ pulses with FWHM $2$ optical cycles and peak intensity $8\times10^{13}\ \text{W/cm}^2$ in each component. The top row is the single-color, $z$-polarized reference. The remaining rows use equal-frequency orthogonal components with relative phases $0$, $\pi/2$, $\pi$, and $3\pi/2$. Other definitions are as in Fig.~\ref{fig:800nmLongAngle}.}
\label{fig:800nmShortAngle}
\end{figure}

For the shorter pulses with FWHM $2$ optical cycles, the angular distributions depend strongly on the relative phase, as shown in Fig.~\ref{fig:800nmShortAngle}. Changing the phase of the $x$-polarized component rotates the preferred emission direction from quadrants in which the $x$ and $z$ components have the same sign to quadrants in which they have opposite signs. The relative phase of the orthogonal field can therefore steer the fragment emission direction.

\section{Conclusions and discussion}
We have presented full-dimensional quantum simulations of $H_2^+$ dissociative ionization in strong orthogonal laser fields. The central result is an additional proton KER peak at approximately $4-5$ eV that is absent from the corresponding single-color, linearly polarized calculations. The perpendicular field enhances excitation of specific molecular eigenstates, particularly $E_2$ and $E_4$, and calculations initiated from these states reproduce the secondary peak. This evidence associates the new channel with laser-induced vibrational excitation.

The relative yield of the secondary peak can be controlled through the phase, frequency, and intensity of the perpendicular field. Orthogonal fields also weaken the simple electron--nuclear energy-sharing pattern observed under linear polarization and steer both proton and electron emission away from the principal $z$ axis. These results demonstrate that the spatial geometry and relative phase of a multidimensional field provide effective control parameters for correlated electron--nuclear fragmentation dynamics.
\section*{Acknowledgments}
The author thanks Prof. Armin Scrinzi for suggesting the investigation of vibrational excitation and for fruitful discussions that helped clarify the mechanism underlying the secondary KER peak.
\appendix
\section{Flat-top pulse shape}
\begin{equation}\label{eq:flatTop}
  a(t)=f_{\frac{n}{2}\tau,(\frac{n}{2}+1)\tau}(-t)f_{\frac{n}{2}\tau,(\frac{n}{2}+1)\tau}(t),-(1+\frac{n}{2})\tau \leq t \leq(1+\frac{n}{2})
\end{equation}
with truncation function
\begin{equation}\label{eq:truncationFunction}
f_{\alpha,\beta}(r):=\left\{\begin{matrix}
1 & r < \alpha\\
\frac{2}{(\alpha-\beta)^3}(r-\beta)^2(r-\frac{3\alpha-\beta}{2}) & \alpha\leq r<\beta  \\ 
0&r\geq\beta 
\end{matrix}\right.
\end{equation}
\section{Joint energy spectra}\label{sec:spectrumShort}

\begin{figure}[H]
\centering
\includegraphics[width=0.23\textwidth,trim=1.2cm 0cm 1.5cm 0.5cm,clip]{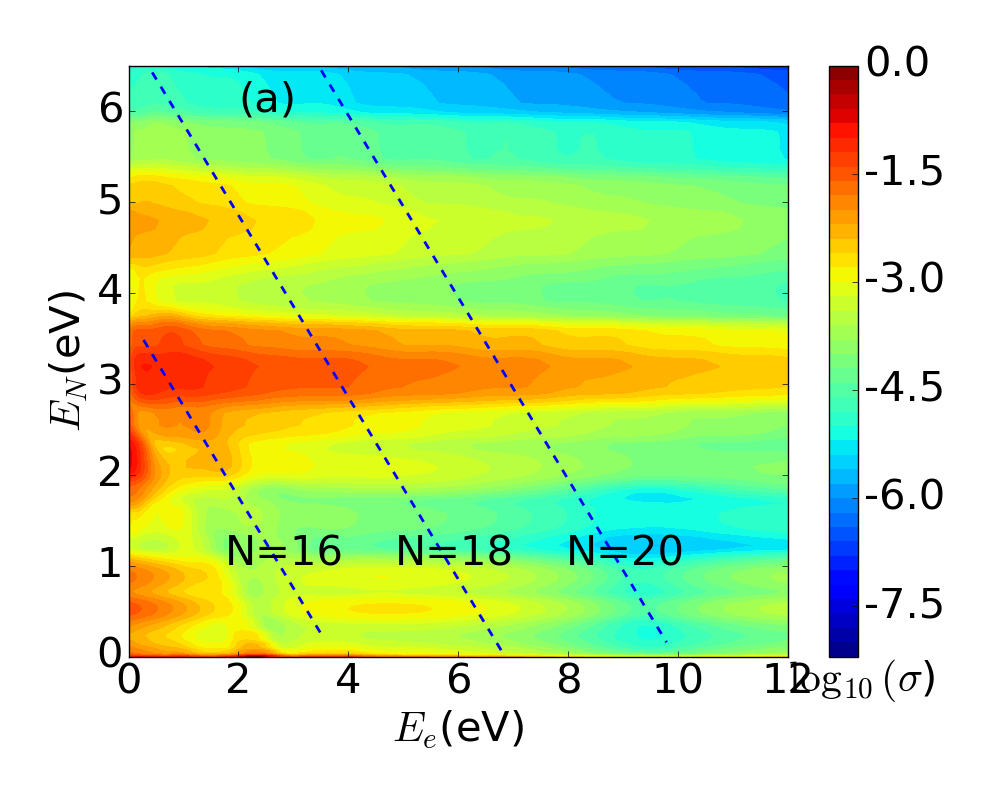}
\includegraphics[width=0.23\textwidth,trim=1.2cm 0cm 1.5cm 0.5cm,clip]{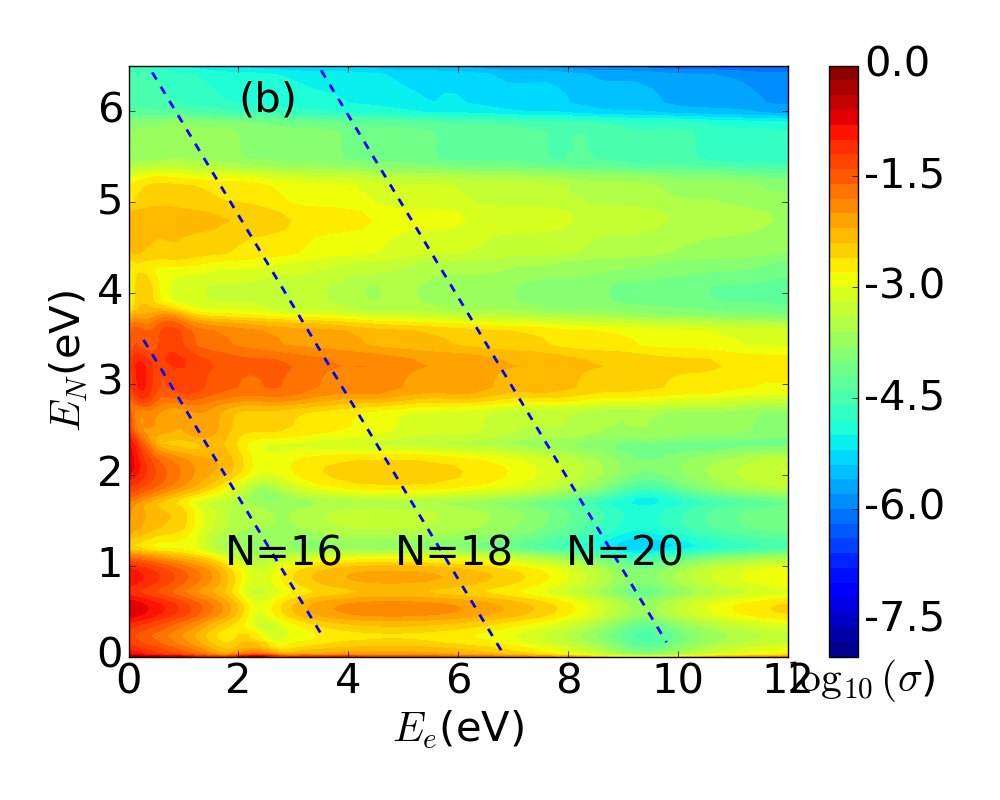}
\includegraphics[width=0.23\textwidth,trim=1.2cm 0cm 1.5cm 0.5cm,clip]{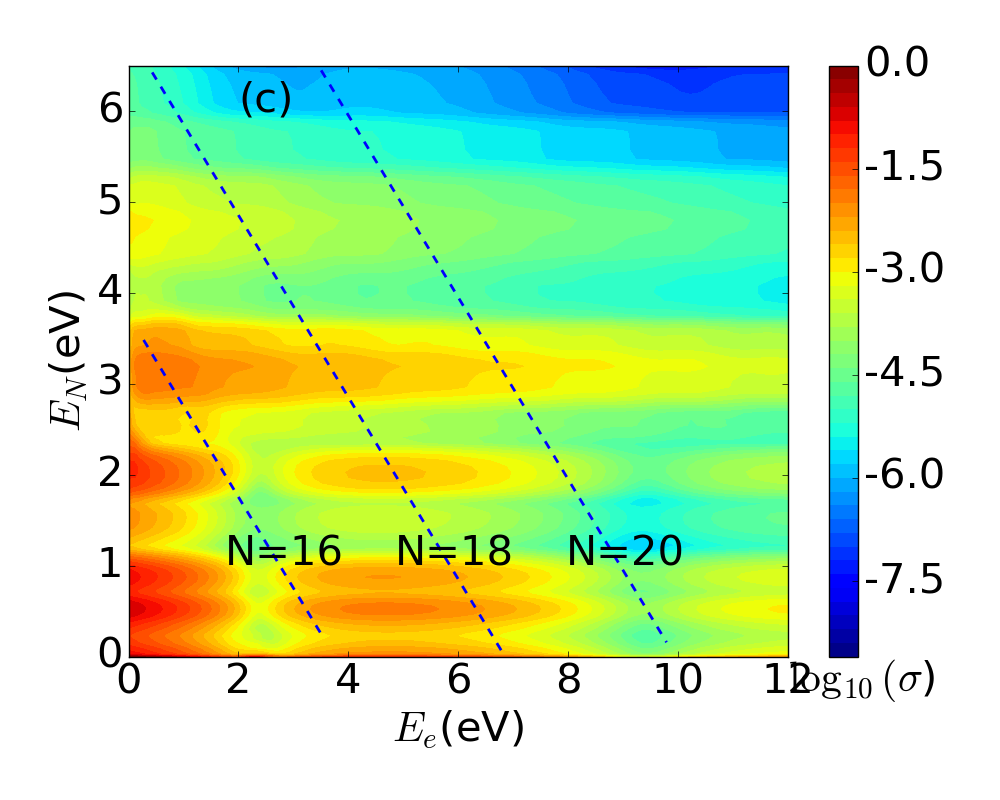}
\includegraphics[width=0.23\textwidth,trim=1.2cm 0cm 1.5cm 0.5cm,clip]{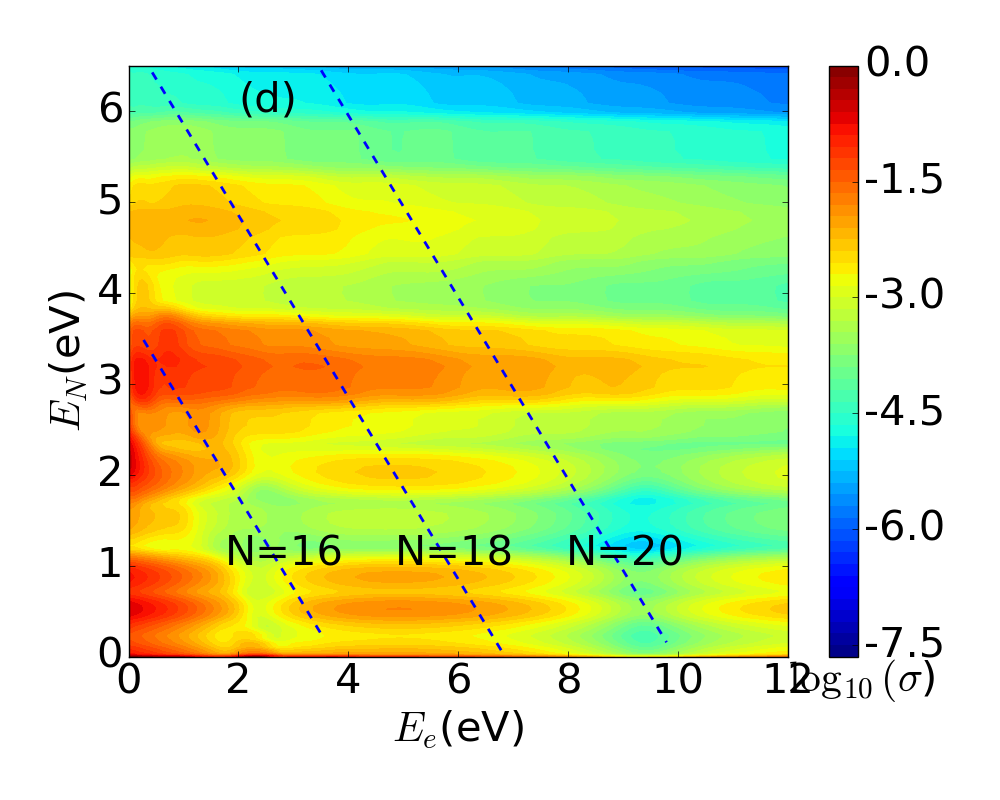}
\caption{JES for orthogonal $800$-nm ($z$-polarized) and $400$-nm ($x$-polarized) $\cos^8$ pulses, each with peak intensity $8\times10^{13}\ \text{W/cm}^2$. The relative phases of the $400$-nm component are (a) $0$, (b) $\pi/2$, (c) $\pi$, and (d) $3\pi/2$. The dashed lines show the energy-sharing condition of Eq.~\ref{eq:energySharing}.}
\label{fig:normal400}
\end{figure}
\begin{figure}[H]
\centering
\includegraphics[width=0.23\textwidth,trim=1.2cm 0cm 1.5cm 0.5cm,clip]{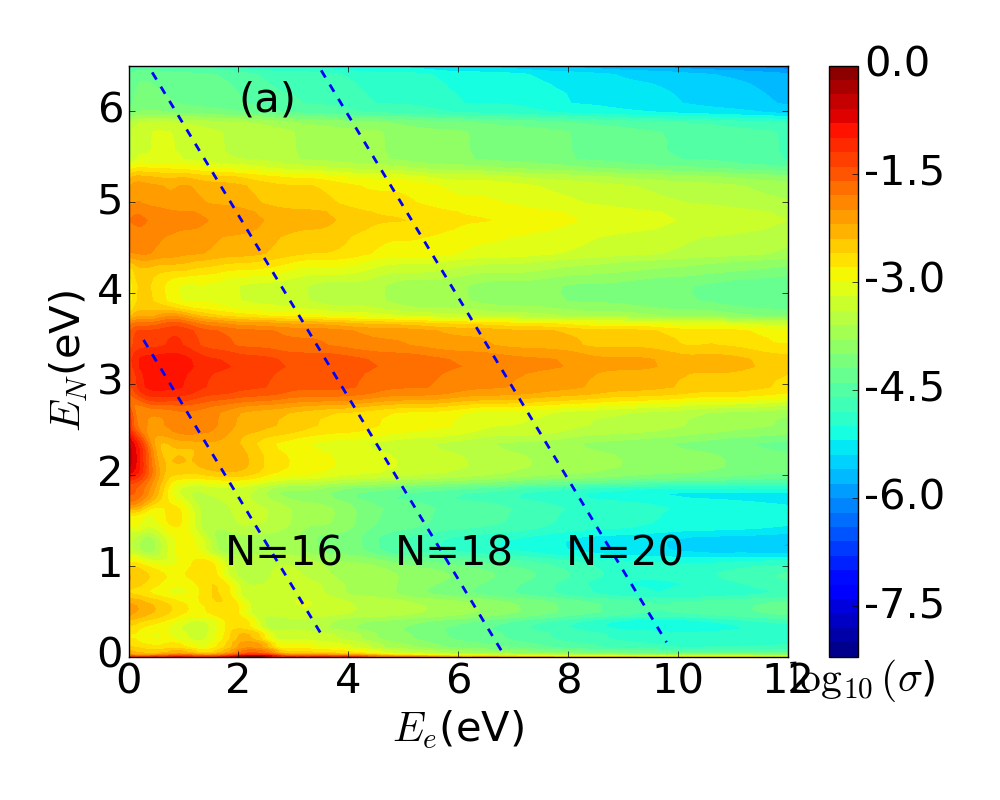}
\includegraphics[width=0.23\textwidth,trim=1.2cm 0cm 1.5cm 0.5cm,clip]{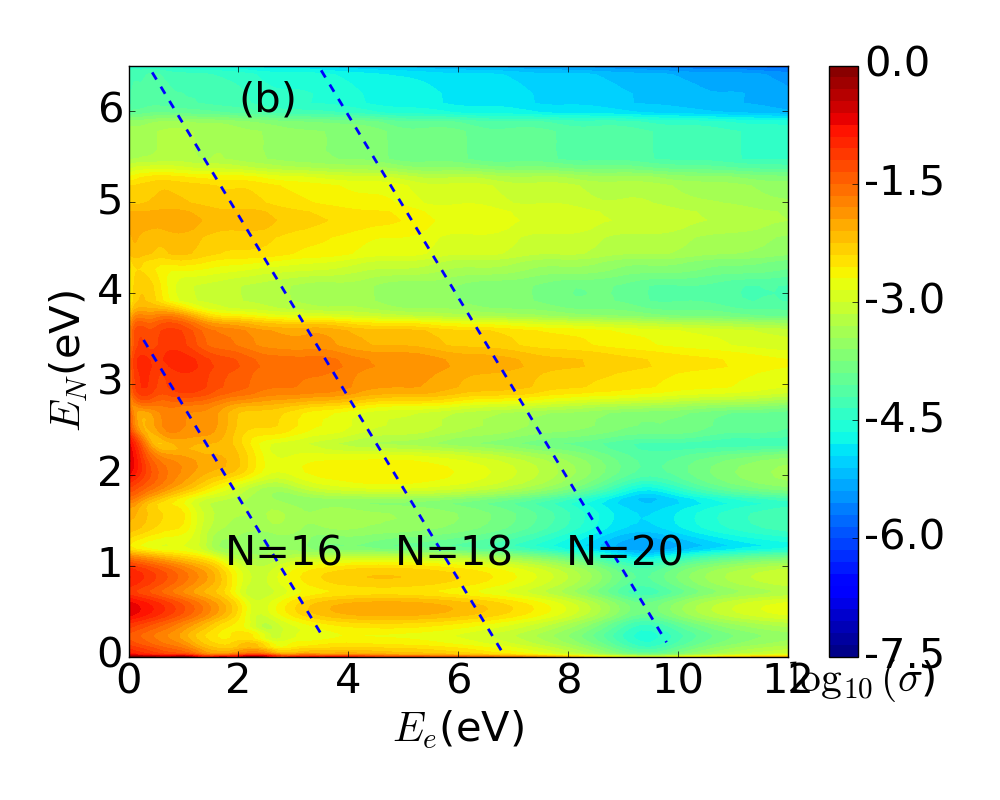}
\includegraphics[width=0.23\textwidth,trim=1.2cm 0cm 1.5cm 0.5cm,clip]{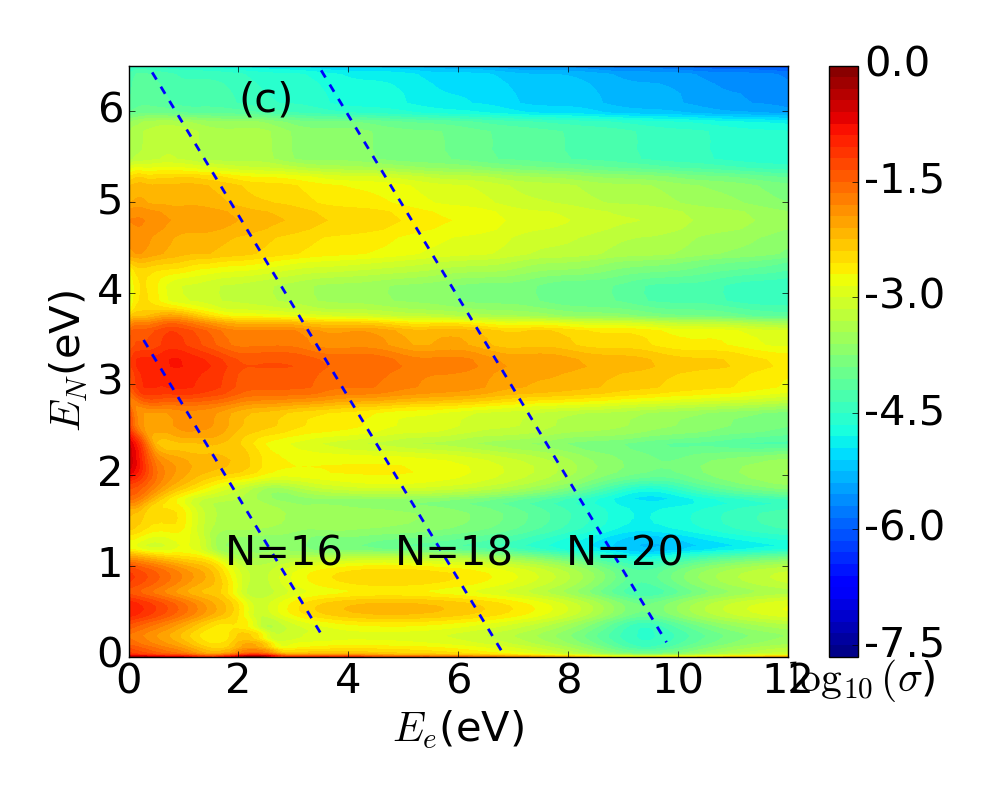}
\includegraphics[width=0.23\textwidth,trim=1.2cm 0cm 1.5cm 0.5cm,clip]{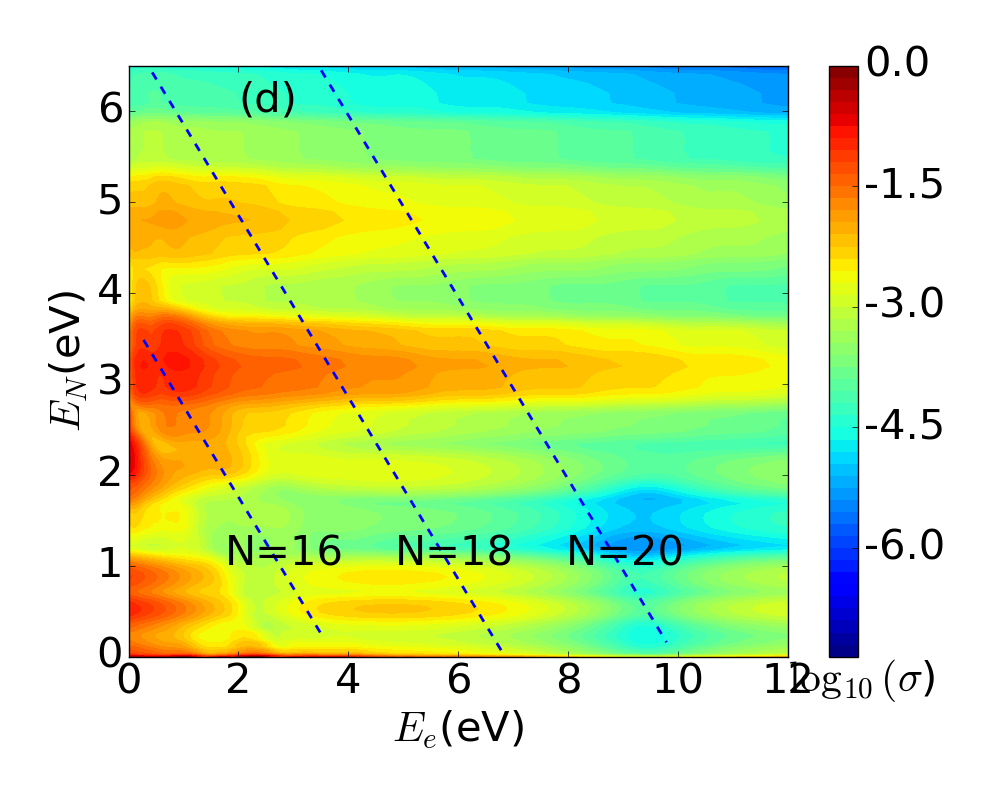}
\caption{JES for an $800$-nm, $z$-polarized pulse with peak intensity $8\times10^{13}\ \text{W/cm}^2$ and an orthogonal $400$-nm, $x$-polarized pulse with peak intensity $1.6\times10^{14}\ \text{W/cm}^2$. Both pulses have $\cos^8$ envelopes. The relative phases of the $400$-nm component are (a) $0$, (b) $\pi/2$, (c) $\pi$, and (d) $3\pi/2$. The dashed lines show the energy-sharing condition of Eq.~\ref{eq:energySharing}.}
\label{fig:high400}
\end{figure}
\begin{figure}[H]
\centering
\includegraphics[width=0.23\textwidth,trim=1.2cm 0cm 1.5cm 0.5cm,clip]{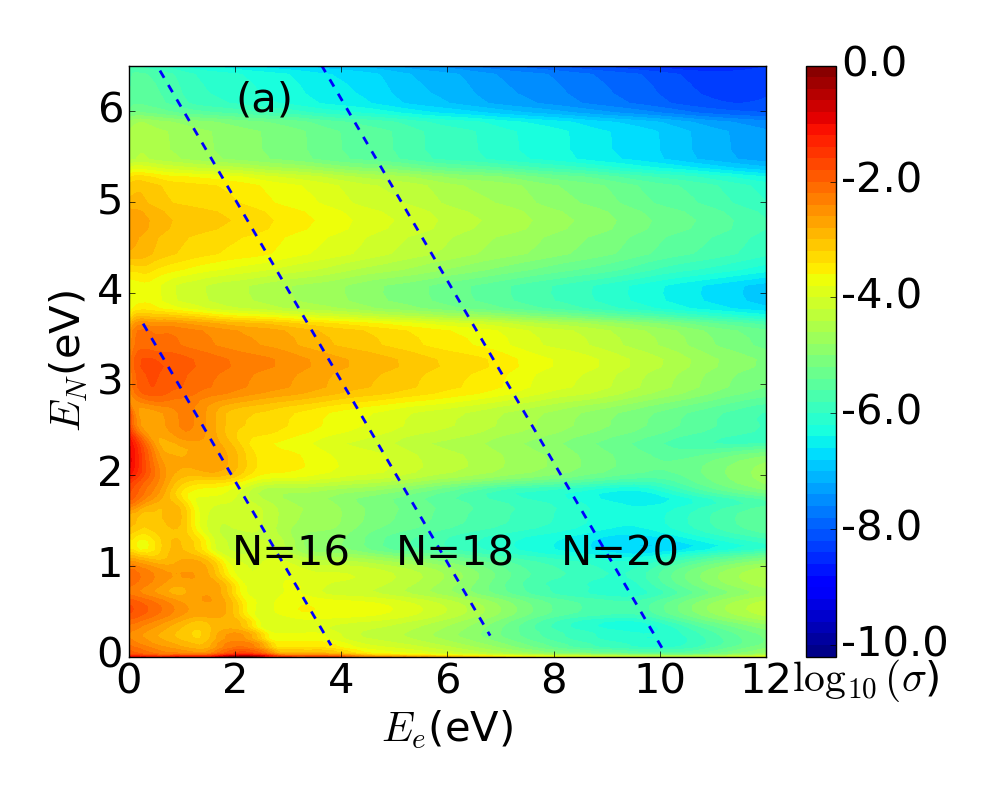}
\includegraphics[width=0.23\textwidth,trim=1.2cm 0cm 1.5cm 0.5cm,clip]{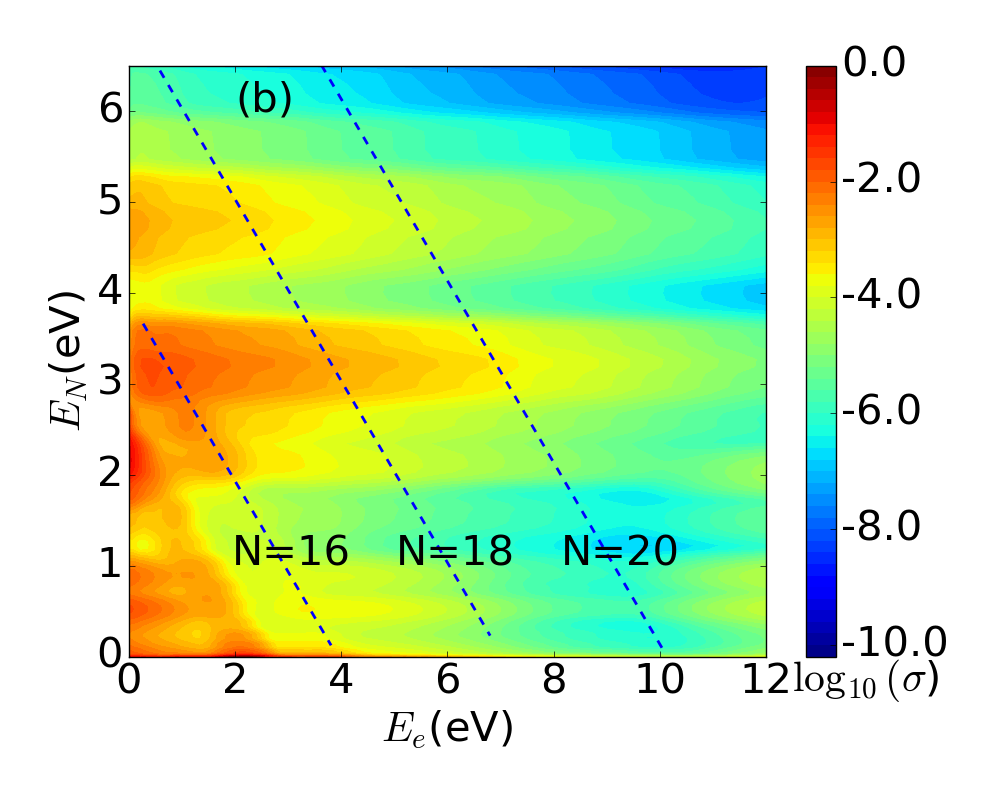}
\includegraphics[width=0.23\textwidth,trim=1.2cm 0cm 1.5cm 0.5cm,clip]{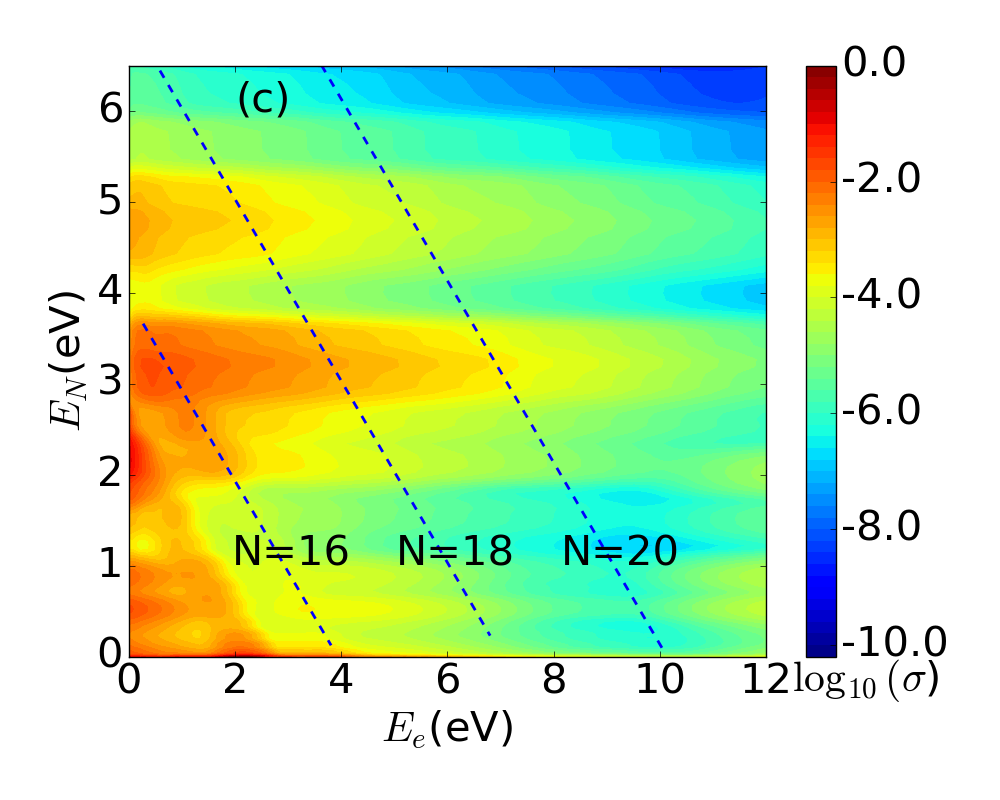}
\includegraphics[width=0.23\textwidth,trim=1.2cm 0cm 1.5cm 0.5cm,clip]{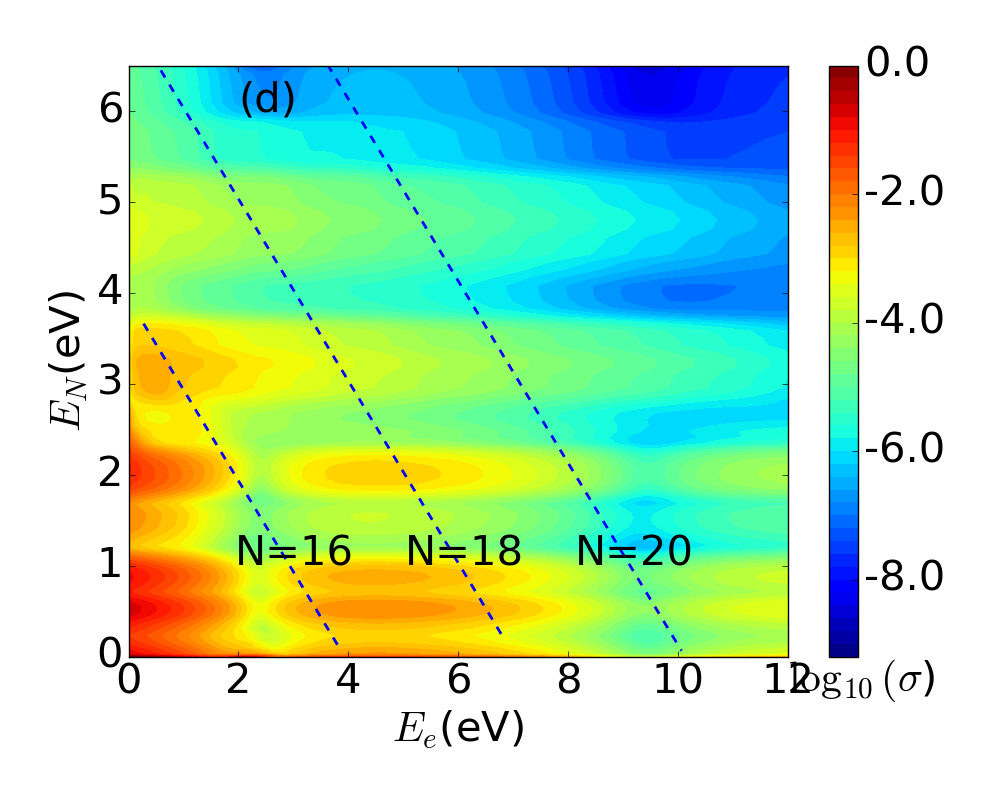}
\caption{JES for orthogonal $z$- and $x$-polarized $800$-nm pulses with $\cos^8$ envelopes and peak intensity $8\times10^{13}\ \text{W/cm}^2$ in each component. The relative phases of the $x$-polarized component are (a) $0$, (b) $\pi/2$, (c) $\pi$, and (d) $3\pi/2$. The dashed lines show the energy-sharing condition of Eq.~\ref{eq:energySharing}.}
\label{fig:800nsShort}
\end{figure}
\bibliography{h2plus.bib}
\end{document}